\documentstyle[12pt]{article}
\def\mn{{\mu\nu}}

\def\frac#1#2{{\textstyle{{#1}\over {#2}}}}

\def\lsim{\mathrel{\rlap{\lower4pt\hbox{\hskip1pt$\sim$}}
    \raise1pt\hbox{$<$}}}
\def\gsim{\mathrel{\rlap{\lower4pt\hbox{\hskip1pt$\sim$}}
    \raise1pt\hbox{$>$}}}
\def\sqr#1#2{{\vcenter{\vbox{\hrule height.#2pt
         \hbox{\vrule width.#2pt height#1pt \kern#1pt
         \vrule width.#2pt}
         \hrule height.#2pt}}}}

\newcommand{\beq}{\begin{equation}}
\newcommand{\eeq}{\end{equation}}
\newcommand{\bea}{\begin{eqnarray}}
\newcommand{\eea}{\end{eqnarray}}

\renewenvironment{thebibliography}[1]
 { \rm
   \begin{list}{\arabic{enumi}.}
    {\usecounter{enumi} \setlength{\parsep}{0pt}
     \setlength{\itemsep}{3pt} \settowidth{\labelwidth}{#1.}
     \sloppy
    }}{\end{list}}
 
\begin{document}
\titlepage
 
\begin{flushright}
{NCUSF 3\\}
{Jan 2010\\}
\end{flushright}

\vglue 1cm
	    
\begin{center}
{{\bf Factoring the Dispersion Relation in the Presence of Lorentz Violation\\}
\vglue 1.0cm
{Don Colladay, Patrick McDonald, and David Mullins\\} 
\bigskip
{\it New College of Florida\\}
\medskip
{\it Sarasota, FL, 34243, U.S.A.\\}
 
\vglue 0.8cm
}
\vglue 0.3cm
 
\end{center}
 
{\rightskip=3pc\leftskip=3pc\noindent
We produce an explicit formula for the dispersion relation for the
Dirac Equation in the Standard Model Extension (SME) in the presence
of Lorentz violation.  Our expression is obtained using a novel
techniques which exploit the algebra of quaternions.  The dispersion
relation is found to conveniently factor in two special cases that
each involve a mutually exclusive set of non-vanishing
Lorentz-violating parameters.  This suggests  that a useful approach
to studies of Lorentz-violating models is to split the parameter space
into two separate pieces, each of which yields a simple, tractable
dispersion relation that can be used for analysis. } 

\vskip 1 cm

\begin{center}
{\it Submitted for Publication in JMP\\}

\end{center}

\newpage
 
\baselineskip=20pt
 
{\bf \noindent I. INTRODUCTION}
\vglue 0.4cm

As a low energy limit of a fundamental theory, the Standard Model (SM)
has proven a remarkable success.  As a result, well-known properties of
the model, in particular, Lorentz invariance, have become default
characteristics built in to many models which purport to underlie the
SM at higher energy scales.  While experimental
evidence indicates that in presently accessible energy regimes violation of
Lorentz symmetry must be small, there is no a priori necessity for
insisting that Lorentz invariance hold for theories whose intent is to
probe physics at higher energy.  Indeed, Lorentz violation may serve as
a signal for new physics at the Plank scale \cite{kps}.  

Motivated by these and related observations, a model has been
constructed which facilitates the study of the possible effects of
incorporating Lorentz violation in the SM.  This model, called the
Standard Model Extension (SME), has been studied extensively \cite{ck,cpt98}.  
The SME framework exhibits many of the properties of standard quantum
field theories including gauge invariance, energy-momentum
conservation, causality and stability (in concordant
frames)\cite{kle}, observer Lorentz invariance and hermiticity. 
In addition, numerous renormalizability properties of the theory have
been established \cite{klp1,klp2,cm1}, etc, and various implications for
particle theory, gravity (Lorentz violation provides an alternative
means of generating theories of gravity \cite{kpott}) and cosmology (see
 \cite{kmewes} for a study of the relationship of Lorentz violation to
 cosmic microwave background data) have been discussed.

In addition to the theoretical work done on the SME, numerous
experiments have been performed to bound the LV effects predicted by
the theory.  These experiments involve numerous aspects of the SM (bounds
associated to electrons, photons, neutrinos, and hadrons, etc)   An
exhaustive list and a brief discussion of relevant
 experimental results are contained in a well-maintained set of data
 tables \cite{neil}. 

Of central importance in the design, implementation
and interpretation of any experiment intended to probe Lorentz
violation is a precise understanding of the associated dispersion
relation.  The goal of this paper is to give a detailed description of
the dispersion relation for the Dirac operator associated to the free fermion
sector of the SME.
In carrying out our goal we discover some new properties of the dispersion relation.
In particular, our analysis provides a natural splitting of the
parameters determining Lorentz violation into two disjoint sets. This
partition includes as special cases all previously investigated
special cases in which a precise description of the dispersion
relation is easy to establish.  In addition, our analysis provides new
results for the special case in which Lorentz violation is coupled to
spin.  While we offer no explanation for the
emergence of our partition as a manifestation of the properties of the
SME, the possibility that hidden symmetries might be their source
provides an intriguing problem for future investigation.

This paper is organized as follows:  In the second section we provide
a introduction to the SME in which we establish both notation and
basic defining features of the model.  In the next section we use a
representation involving the quaternions to
investigate the Dirac equation and associated dispersion relation for
the SME.  In this section we also establish our fundamental result: an
explicit polynomial representation for the dispersion relation in
terms of the Lorentz violating parameters which define the theory.  In
the fourth section we discuss consequences of our formula, including
consistency checks with existing literature, a complete analysis of
several important special cases, and the discovery of new
relationships between Lorentz violating parameters.  In the final
section of the paper we discuss our conclusions and some potential directions for
future work.

\vglue 0.6cm
{\bf \noindent II. NOTATION, CONVENTIONS AND BACKGROUND}
\vglue 0.4cm

Let $\sigma_j$ be the Pauli matrices, $\gamma_\mu$ denote the standard
gamma matrices, and $\eta_{\lambda \mu} = {\rm diag}(1,-1,-1,-1)$ the Minkowski metric with
signature -2.  Thus,
\beq
i\sigma_1  =  \left( \begin{matrix}{
  0 & i \cr
  i & 0 }
\end{matrix} \right) \ \ \ \ \hspace{.25in} 
i\sigma_2 = \left( \begin{matrix}{
  0 & 1 \cr
  -1 & 0 }
\end{matrix} \right) \ \ \ \ \hspace{.25in} 
i\sigma_3 = \left(\begin{matrix}{
  i & 0 \cr
  0 & -i }
\end{matrix}\right) 
\eeq
and 
\beq
\gamma^0 =  \left( \begin{matrix}{
  I & 0 \cr
  0 & -I }
\end{matrix} \right)  \ \ \ \ \hspace{.25in} \gamma^j =
\left( \begin{matrix} {
  0 & \sigma_j \cr
  -\sigma_j & 0 }
\end{matrix} \right)  \label{gamma2.1}
\eeq
where $i$ is the imaginary unit. The SME Lorentz violating Lagrangian
for a single spin-$\frac{1}{2}$ fermion is given by
\beq
{\mathcal L} =  i \bar{\psi}\Gamma^\nu \partial_\nu \psi -
\bar{\psi}M \psi \label{lag2.1}
\eeq
where 
\bea
\Gamma^\nu & = & \gamma^\nu + c^{\mu\nu}\gamma_\mu + d^{\mu\nu}\gamma_5 \gamma_\mu
+ e^\nu + if^{\nu}\gamma_5 + \frac12 g^{\lambda \mu \nu}
\sigma_{\lambda \mu}, \label{Gamma2.1} \\
M & = & m + a_\mu\gamma^\mu + b_\mu\gamma_5 \gamma^\mu + \frac12
H_{\mu\nu}\sigma^{\mu\nu}\label{mass2.1}.
\eea

The coefficients $c_{\mu\nu}, \ d_{\mu \nu}, \ e_\nu, \ f_\nu,
\ g_{\lambda \mu \nu}, \ a_\mu, \ b_\mu, \ H_{\mu\nu}$ governing
Lorentz violation are assumed small.  Since the Lagrangian is
Hermitian, the parameters are real.  In addition, the parameters 
$c_{\nu\mu}$ and $d_{\nu\mu}$ can be taken to be traceless, $H_{\mu\nu}$ 
  antisymmetric, and $g^{\lambda\mu\nu}$ antisymmetric in the first
  two components.  The parameters $a_{\mu}, \ b_\mu, \ H_{\mu\nu},$
  have the dimension of mass, while the remaining parameters are
  dimensionless. 
  
As mentioned above, the SME exhibits many of the properties of
standard quantum field theories including gauge invariance,
energy-momentum conservation, causality, stability, observer Lorentz
invariance, hermiticity and power counting renormalizability.  In
addition, any theory that generates the SM and exhibits spontaneous
Lorentz and CPT violation contains the SME as an appropriate limit
\cite{ck}.

The Dirac equation associated to the Lagrangian (\ref{lag2.1}) is
given by  
\beq
  (i\Gamma^\nu \partial_\nu - M)\psi =0 
\eeq
or, in momentum space coordinates (using $\psi(x) = e^{- i p \cdot x} u(p)$ for now)
\beq
  (\Gamma^\nu p_\nu - M)\psi =0. \label{dirac2.1}
\eeq
The Dirac operator $(\Gamma^\nu p_\nu - M)$ is a $4\times 4$ matrix with
complex entries.  The dispersion relation characterizes the null space
of the Dirac operator and is given by  
\beq
  {\rm det} (\Gamma^\nu p_\nu - M) =0. \label{dispersion2.1}
\eeq

Expression (\ref{dispersion2.1}) describes the zeroes of a fourth
order polynomial in $p^0$ whose coefficients depend smoothly on the
Lorentz violating parameters and on the momentum vector
$\vec{p}=(p^1,p^2,p^3).$  The explicit covariant form of this dispersion relation is
presented in the literature \cite{ralph2},
however, the general expression of the dispersion relation yields little insight
into specific parameter choices that allow for a simple factorization.
The present work remedies the situation by using new techniques to analyze
the resulting algebraic varieties that arise from this dispersion relation. 

We will denote by ${\mathcal C}$ the charge conjugation operator,
${\mathcal P}$ the parity inversion operator, and ${\mathcal T}$ the
time reversal operator.    It is of interest to
understand the effect of these discrete symmetries on the structure of the
Lorentz-violating theory.

Using the Dirac representation we write the charge conjugation matrix
as 
\beq
C = i\gamma^2\gamma^0 = \left(\begin{matrix}{
 0 &  -i\sigma_2  \cr  -i\sigma_2 & 0 }\end{matrix} \right).\label{chargecon2.1}
\eeq

The C, P and T symmetries of
the SM are given at the level of the Dirac operator by the transformations 
\begin{eqnarray}
{\mathcal C}:  \Gamma^\nu p_\nu - M & \longrightarrow & C\gamma^0 (\Gamma^\nu p_\nu -
M)^* C \gamma^0  \label{Csym2.1} \\
{\mathcal P}:  \Gamma^\nu p_\nu - M & \longrightarrow &  \gamma^0 (\Gamma^\nu p_\nu -
M)\gamma^0  \label{Psym2.1}\\
{\mathcal T}:  \Gamma^\nu p_\nu - M & \longrightarrow &  i\gamma^5 C(\Gamma^\nu p_\nu
- M)^* i\gamma^5 C .\label{Tsym2.1}
\end{eqnarray}

From (\ref{dispersion2.1}) and basic invariant theory of the real valued determinant, it is clear
that the dispersion relation is unaltered by any of the above transformations.  
This indicates that the generic effect of these discrete transformations is to permute the energy
eigenstates in various ways.
The effect of CPT on
the SME Lagrangian can be neatly summarized for our purposes.  For
fixed $\vec{p}, \ b^\mu, \ d^{\mu \nu},\  g^{\mu \nu \lambda}, \  H^{\mu \nu}$
we will denote by $Rt(p^0(\vec{p}, b^\mu, d^{\mu \nu}, g^{\mu \nu
  \lambda} H^{\mu \nu}))$ the roots of the fourth order polynomial in
$p_0$ defined by (\ref{dispersion2.1}).  Then the above transformation properties of
the Dirac operator yield the following root permutations:
\begin{eqnarray*}
{\mathcal C}: Rt(p^0(\vec{p}, b^\mu, d^{\mu \nu}, g^{\mu \nu
  \lambda}, H^{\mu \nu}))  & \longrightarrow & -Rt(p^0(-\vec{p}, b^\mu,
-d^{\mu \nu}, g^{\mu \nu   \lambda}, -H^{\mu \nu}))    \\
{\mathcal P}:  Rt(p^0(\vec{p}, b^\mu, d^{\mu \nu}, g^{\mu \nu
  \lambda}, H^{\mu \nu}))  & \longrightarrow & Rt(p^0(-\vec{p}, -b_\mu,
-d_{\mu \nu}, g_{\mu \nu   \lambda}, H_{\mu \nu}))    \\
{\mathcal T}: Rt(p^0(\vec{p}, b^\mu, d^{\mu \nu}, g^{\mu \nu
  \lambda}, H^{\mu \nu}))  & \longrightarrow & Rt(p^0(-\vec{p}, b_\mu,
d_{\mu \nu}, -g_{\mu \nu   \lambda}, -H_{\mu \nu}))    
\end{eqnarray*}

\vglue 0.6cm
{\bf \noindent III. DISPERSION RELATION USING QUATERNIONS}
\vglue 0.4cm

To further analyze the dispersion relation (\ref{dispersion2.1}) we
employ a quaternion algebra.  More precisely, let $1,
\ \hat{i},\ \hat{j},\ \hat{k}$ denote the usual basis for the
quaternions (denoted ${\bf H}$), and let $i$ denote the usual
complex imaginary unit.  Elements of ${\bf H}$ are
expressions of the form $a+\hat{i}b +\hat{j}c + \hat{k}d$ where $a,
\ b, \ c, \ d$ are real.  Denote by a ``bar'' quaternionic
conjugation:  
\beq
\bar{\hat{i}} = -\hat{i} \hspace{.25in} \bar{\hat{j}} =
-\hat{j} \hspace{.25in} \bar{\hat{k}} = -\hat{k} \hspace{.25in}
\bar{i} = i. \hspace{.25in} 
\eeq
We can identify quaternions with $2\times 2$ complex matrices of the
form $\left( \begin{matrix}{
  z  & w \cr
  -\bar{w} & \bar{z} }
\end{matrix} \right): $ 
\begin{equation}\label{iso2.1}
a+\hat{i}b +\hat{j}c + \hat{k}d \rightarrow \left( \begin{matrix}{
  a+ib  & c+id \cr
  -c+id & a-ib }
\end{matrix} \right).
\end{equation}
This identification is an isomorphism of skew fields; in particular
multiplication of quaternions is mapped to matrix multiplication.
Note that the map identifies the Pauli matrices with elements of the
standard basis of the quaternions:
\beq
i\sigma_1  \longrightarrow  \hat{k} \hspace{.25in} i\sigma_2
\longrightarrow  \hat{j}  \hspace{.25in} i\sigma_3 \longrightarrow
\hat{i} .
\eeq
We will refer to matrices of the form (\ref{iso2.1}) as {\it
  quaternionic matrices.}  These matrices play a central role in our
approach\footnote{Our approach should not be confused with attempts in
  the literature to reformulate quantum mechanics itself in terms of
  quaternions \cite{quatdirac}.  We simply use the algebraic properties
  of the quaternions as a computational tool in the conventional
  framework of complex-valued fields.}.  

Quaternionic matrices can be used to represent general $2\times 2$
complex matrices.  More precisely, every $2\times 2$ complex matrix
has a unique representation of the form  $A+iB$ where $A$ and $B$ 
are quaternionic.  This representation extends inductively to higher
dimension.  In particular, given a general $4\times 4$ complex matrix
$M,$ we can decompose $M$ into four $2\times 2$ complex matrices.
Since each complex $2\times 2$ block has a unique representation of
the form $A+iB$ with $A$ and $B$ quaternionic, $M$ has a unique
decomposition of the form $M_1+iM_2$ where each $M_i$ is a $2\times 2$
block matrix with each block given by a quaternionic matrix.  Since
each block is quaternionic, every $4\times 4$ complex matrix has a
unique representation of the form $Q_1+iQ_2$ where the $Q_i$ are
$2\times 2$ matrices with entries which are quaternions.  Implementing
this construction for the gamma matrices gives
\beq
\gamma^1  \longrightarrow  i \left( \begin{matrix}{
  0  & -\hat{k} \cr
  \hat{k} & 0 }
\end{matrix} \right) ,\hspace{.25in} \gamma^2
\longrightarrow   i \left( \begin{matrix}{
  0  & -\hat{j} \cr
  \hat{j} & 0 }
\end{matrix} \right) , \hspace{.25in} \gamma^3 \longrightarrow
 i \left( \begin{matrix}{
  0  & -\hat{i} \cr
  \hat{i} & 0 }
\end{matrix} \right),
\eeq
and the standard Dirac equation becomes the $2\times 2$ matrix
equation 
\beq
\left[ \left( \begin{matrix}{
  p_0-m  & 0 \cr
  0 & -p_0-m }
\end{matrix} \right) +i \left( \begin{matrix}{
  0 & \hat{p} \cr
  - \hat{p} & 0 }
\end{matrix}\right) \right] \left(\begin{matrix}{
  \phi \cr
  \xi }
\end{matrix}\right) = \left( \begin{matrix}{ 0 \cr
0}
  \end{matrix} \right) \label{nonperturbed2.1},
\eeq
where 
\beq
\hat{p}  = i \vec p \cdot \vec \sigma = p_3\hat{i} +p_2\hat{j} + p_1\hat{k}.
\eeq
We will henceforth always use a ``hat'' to represent quaternions
with no real component and standard script to denote quaternions with
no non-real component.  Note the reversal of first and third components of
the imaginary quaternion relative to the standard three-vector
notation.  With this convention we can express the Dirac equation for
the SME Lagrangian:

\beq
\left[ \left(\begin{matrix}{
  x  & \alpha_0+\hat{\alpha} \cr
  -\alpha_0 +\hat{\alpha} & -y }
\end{matrix}\right)  +i \left(\begin{matrix}{
  \hat{\epsilon} & \hat{p} \cr
  -\hat{p} & \hat{\delta} }
\end{matrix}\right) \right] \left(\begin{matrix}{
  \phi \cr
  \xi }
\end{matrix}\right) = \left( \begin{matrix} {
0 \cr
0  }
\end{matrix} \right) \label{dirac2.2}
\eeq
where $x=p_0-m,$ $y=p_0+m,$ and $\hat{\alpha}, \ \hat{\delta},
\ \hat{\epsilon}$ are quaternions with no real part.  The relationship
between the quaternionic parameters and the SME Lorentz violating
parameters can be made explicit.  Let $\epsilon^{ijk}$ be the totally
antisymmetric symbol on three letters and set 
\begin{eqnarray*}
\hat{d}^i_1   =  d^{0i}  & \hspace{.25in}& 
 \hat{d}^i_p  =  d^{ij}p_j  \\ 
\hat{H}^i   =  H^{0i}  & \hspace{.25in}& 
\hat{G}^i  =  g^{0ij}p_j  \\ 
\hat{h}^i   =  1/2 \epsilon^{ijk}H^{jk}  & \hspace{.25in}& 
\hat{g}^i  =  1/2 \epsilon^{ijk}g^{jkl}p_l.
\end{eqnarray*}
Then 
\begin{eqnarray}
\alpha_0  =  b^0 + \hat{d}_1\cdot \hat{p}   & \hspace{.25in}& \hat{\alpha}  = 
\hat{H} - \hat{G} \label{alphadef3.1} \\
\hat{\epsilon}  =  \hat{b} +\hat{d}_p +(\hat{g}-\hat{h}) & \hspace{.25in} & 
\hat{\delta}  =  -\hat{b} -\hat{d}_p +
(\hat{g}-\hat{h}). \label{epsilondef3.1}  
\end{eqnarray}
Note that $a^\mu$, $e^\mu$,  and $c^{\mu\nu}$ do not appear in the
above expressions.  These parameters can in fact all be absorbed into
a redefined set of momenta and mass parameters that are simply related
to the physical momenta and masses. 
\begin{eqnarray}
\hat p^\prime & = & \hat p - \hat a - \hat c_p \\
p_0^\prime &=& (1 + c^{00}) p_0 - a_0 - \vec c_1 \cdot \vec p \\ 
m^\prime& =& m - \vec e \cdot \vec p \quad ,
\end{eqnarray}
where  $(c_p)^i = c^{ij} p^j$ and $(c_1)^i = c^{0i}$.
The primes on these parameters are dropped from the following
calculation for notational convenience since they do not change the
general quaternionic structure of the equations.  In addition, several of
these parameters may be removed using appropriate field redefinitions
and are therefore not physical anyway \cite{colmac2}.
At the end of the
calculation, these parameters may be easily included by inverting the
above linear equations and solving for the physical momenta and mass
parameters.  Note that the results of \cite{ba} and \cite{ralf}
have been used to remove the $f_\mu$ dependence and to arrange for $\Gamma^0 =
\gamma^0.$  These reductions greatly simplify the  computations to
follow.

We can use (\ref{chargecon2.1})-(\ref{Tsym2.1}) to compute the effect
of ${\mathcal C}, \ {\mathcal P},$ and ${\mathcal T}$ on the Dirac
equation in our representation.  For charge conjugation we have
\begin{equation}
{\mathcal C}:  \left(\begin{matrix}{
  x  & \alpha_0+\hat{\alpha} \cr
  -\alpha_0 +\hat{\alpha} & -y }
\end{matrix}\right)  +i \left(\begin{matrix}{
  \hat{\epsilon} & \hat{p} \cr
  -\hat{p} & \hat{\delta} }
\end{matrix}\right)   \longrightarrow   \left(\begin{matrix}{
  -y  & \alpha_0+ -\hat{\alpha} \cr
  -\alpha_0 -\hat{\alpha} & x }
\end{matrix}\right)  +i \left(\begin{matrix}{
  -\hat{\delta} & -\hat{p} \cr
  \hat{p} & -\hat{\epsilon}}
\end{matrix}\right) ,
\end{equation}
while for parity we have
\begin{equation}
{\mathcal P}:  \left(\begin{matrix}{
  x  & \alpha_0+\hat{\alpha} \cr
  -\alpha_0 +\hat{\alpha} & -y }
\end{matrix}\right)  +i \left(\begin{matrix}{
  \hat{\epsilon} & \hat{p} \cr
  -\hat{p} & \hat{\delta} }
\end{matrix}\right)   \longrightarrow 
  \left(\begin{matrix}{
  x  & -\alpha_0- \hat{\alpha} \cr
  \alpha_0 -\hat{\alpha} & -y }
\end{matrix}\right)  +i \left(\begin{matrix}{
  \hat{\epsilon} & -\hat{p} \cr
  \hat{p} & -\hat{\delta} }
\end{matrix}\right) .
\end{equation}
Using the matrix identification $\gamma^5C = -\hat{j}I,$ we can compute
the effect of time reversal:
\begin{equation}
{\mathcal T}: \left(\begin{matrix}{
  x  & \alpha_0+\hat{\alpha} \cr
  -\alpha_0 +\hat{\alpha} & -y }
\end{matrix}\right)  +i \left(\begin{matrix}{
  \hat{\epsilon} & \hat{p} \cr
  -\hat{p} & \hat{\delta} }
\end{matrix}\right)   \longrightarrow 
  \left(\begin{matrix}{ 
  x  & \alpha_0+ \hat{\alpha} \cr
  -\alpha_0 +\hat{\alpha} & -y }
\end{matrix}\right)  +i \left(\begin{matrix}{
  -\hat{\epsilon} & -\hat{p} \cr
  \hat{p} & -\hat{\delta} }
\end{matrix}\right). 
\end{equation}
Using the quaternionic representation of the Dirac equation 
(\ref{dirac2.2}), we can produce a useful expression for the dispersion
relation.  To do so, solve for the spinor $\phi$ in terms of $\xi$ to
obtain  
\beq
\left[(\alpha_0 - \hat{\alpha} +
i\hat{p}) (x-i\hat{\epsilon})(\alpha_0 + \hat{\alpha} +
i\hat{p}) + r(-y+i\hat{\delta})\right] \xi = 0
\eeq
where $r= x^2 -|\hat{\epsilon}|^2.$  We write this as 
\beq
(q_1 +i\hat q_2)\xi = 0 \label{dirac2.3}
\eeq
where 
\begin{eqnarray*}
q_1 & = & x(\alpha_0^2 +\vec{\alpha}^2 +\vec{p}^2) - 2\alpha_0
\vec{\epsilon} \cdot \vec{p} - 2\vec{\alpha} \cdot
(\vec{\epsilon}\times \vec{p}) -ry   \\
\hat q_2 & = & 2x[\alpha_0\hat{p} - (\hat{{p} \times {\alpha}})]
-2\alpha_0 (\hat{{\alpha}\times {\epsilon}})  -2(\vec{\alpha}\cdot
\vec{\epsilon})\hat{\alpha} -2(\vec{p}\cdot
\vec{\epsilon})\hat{p} + ( \vec{\alpha}^2 -\alpha_0^2
+\vec{p}^2)\hat{\epsilon} +r\hat{\delta}    
\end{eqnarray*}
where $\hat{{v}\times {u}}$ represents the
pure imaginary quaternion obtained by forming the cross product of
the two vectors involved and mapping it to the coresponding
quaternion.  

The analysis of (\ref{dirac2.3}) requires the
characterization of the null space of $A+iB$ where $A$ and $B$ are
quaternionic, with $A = aI,$ $a$ real.  It is easy to see that such
operators have null space if and only if $B$ has no real part (ie $B$
is traceless) and $\det(A) = \det(B).$  This latter condition can be
expressed in terms of the quaternionic norm: $\det(A) = \det(B)$ if
and only if $|A|^2 = |B|^2.$  

Returning to (\ref{dirac2.3}), note that $q_1$ is real and that $\hat
q_2$ is a pure imaginary quaternion.  We interpret $i\hat q_2$ as a
linear operator on spinors and (\ref{dirac2.3}) as an eigenvalue
equation for the relevant operator $i\hat q_2.$  By the above
discussion concerning null spaces for quaternionic matrices, the
associated polynomial whose roots determine the desired eigenvalues is
characterized by the equality $|\hat q_2|^2 = q_1^2.$  Explicit
calculation of both sides of this equation yield the dispersion
relation as given by the polynomial identity  
\begin{eqnarray}
\sum_{j=0}^4 \beta_jp_0^j & = &0 \label{disprel}
\end{eqnarray}
where the coefficients satisfy $\beta_4 =1,$ $\beta_3=0,$ and 
\begin{eqnarray}
\beta_2 & = & -2 \left[ \alpha_0^2 +\vec{\alpha}^2 + m^2  + \vec{p}^2
  \right] -( \vec{\epsilon}^2 + \vec{\delta}^2 )  \\ 
\beta_1 & =  &  2m(\vec{\delta}^2 - \vec{\epsilon}^2) - 4\alpha_0 (\vec{p}\cdot
  (\vec{\delta} - \vec{\epsilon})) - 4(\vec{\alpha} \times \vec{p}) \cdot
  (\vec{\epsilon} + \vec{\delta}) \label{lincoeff}  \\ \nonumber
\beta_0 & = & \vec{p}^4  -2\vec{p}^2 ( \alpha_0^2
  +\vec{\alpha}^2 - m^2 + \vec{\epsilon}\cdot \vec{\delta} ) +
  (\alpha_0^2 + \vec{\alpha}^2 +m^2)^2
  - m^2(\vec{\delta}^2 +
\vec{\epsilon}^2) +
2(\vec{\epsilon}\cdot \vec{\delta})(\alpha_0^2 - \vec{\alpha}^2) \\ \nonumber
& & 
+ \vec{\epsilon}^2 \vec{\delta}^2 
+ 4 m[(\vec \alpha \times \vec p) \cdot (\vec \delta - \vec \epsilon) + 
\alpha_0 \vec p \cdot (\vec \delta + \vec \epsilon)]
+ 4 \alpha_0 \vec \epsilon \cdot (\vec \delta \times \vec \alpha) \\ \nonumber
& & 
+   4\left[(\vec \alpha \cdot \vec p)^2  +(\vec \epsilon \cdot \vec p)(\vec p \cdot \vec \delta)  +
(\vec \alpha \cdot \vec \delta)(\vec \epsilon \cdot \vec \alpha)\right] .
\label{constantcoeff} \\  
\end{eqnarray}

The cubic term of the above fourth order monic polynomial vanishes,
therefore the polynomial admits a generic factorization of the form 
\beq
(p_0^2 +tp_0 +u) (p_0^2 -tp_0 +v),
\label{genform3.1}
\eeq
where $t$, $u$, and $v$ are parameters that depend on the vector momentum
as well as the Lorentz-violating parameters, but in general their
explicit expressions are cumbersome and not particularly insightful. 
Note that when the linear term also vanishes, the complete
factorization is particularly simple.  We pursue this and other
consequences of the dispersion formula (\ref{disprel})  in
the next section.

\vglue 0.6cm
{\bf \noindent IV. SPECIAL CASES OF DISPERSION FORMULA}
\vglue 0.4cm

When the Lorentz violating parameters are set to zero, the dispersion 
relation factors into the square of a second-order polynomial
\begin{equation}
(p_0^2 - \vec p ^2 - m^2)^2 = 0.
\end{equation}
Fixing the value of $p_0$ defines a doubly degenerate sphere in momentum space.  
Thus, when the Lorentz violating
parameters are small with respect to mass and momentum, the dispersion
relation is expected to define a smoothly perturbed sphere with possible multiple
sheets.
We show below that the resulting surface is generically a double-sheeted perturbation of
the sphere.

We begin by noticing that the dispersion relation becomes particularly
simple to analyze when the coefficient of the linear term $\beta_1$
given in (\ref{lincoeff}) vanishes.  In order for this to occur for
arbitrary $\vec{p}$ we must have either $\vec{\epsilon} +\vec{\delta}
= 0$ or $\vec{\epsilon} -\vec{\delta} = 0. $  We analyze each case
separately.

{\bf  Case:  $\vec{\epsilon} + \vec{\delta} = 0$} 

When $\vec{\epsilon} = -\vec{\delta}$, to ensure that $\beta_1 = 0$ for arbitrary $\vec p$, it
is sufficient that $\alpha_0=0.$  In this case, the coefficients of the
polynomial (\ref{disprel}) become
\begin{eqnarray}
\beta_2 & = & -2(\vec p^2 + m^2 + \vec \alpha^2 + \vec \delta^2) \\
\beta_0 & = &  (\vec p^2 + m^2 - \vec \alpha^2 + \vec \delta^2)^2 \\
& + & 4\left[\vec \delta^2 \vec \alpha^2  - m^2 \vec \delta^2 + m^2 \vec \alpha^2 + 
(\vec \alpha \cdot \vec p)^2 - (\vec \delta \cdot \vec p)^2 - (\vec \alpha \cdot \vec \delta)^2
+ 2 m (\vec \alpha \times \vec p )\cdot \vec \delta)\right] .\nonumber \\
\end{eqnarray}
 When this is the case, completing the square in the associated dispersion relation
yields
\begin{equation}
p_0^2 =   \vec p^2 + m^2 + \vec \alpha^2+ \vec \delta^2 \pm 2 \sqrt{D_1(\vec p)},
\end{equation}
where 
\beq
D_1(\vec p) = (\vec \alpha \times \vec p - m \vec \delta)^2 + (\vec \delta \cdot \vec p)^2 + 
(\vec \alpha \cdot \vec \delta)^2,
\eeq
is a non-negative quantity, as is required for reality of the energy eigenvalues.

{\bf  Case:  $\vec{\epsilon} - \vec{\delta} = 0$} 

When $\vec{\epsilon} - \vec{\delta}=0$, to ensure that $\beta_1 = 0$
for arbitrary $\vec{p},$ it is sufficient that
$\vec{\alpha}=0.$   The coefficients of the polynomial (\ref{disprel}) become
\begin{eqnarray*}
\beta_2 & = & -2(  \vec{p}^2 + m^2 + \alpha_0^2
 + \vec \delta^2) \\
\beta_0 & = &  \left(\vec p^2 + m^2 - \alpha_0^2 
- \vec \delta^2\right)^2 + 4 m^2 \alpha_0^2 + 8 m \alpha_0 
\vec p \cdot \vec \delta + 4 (\vec p \cdot \vec \delta)^2.
\end{eqnarray*}
Completing the square  yields the solutions
\begin{equation}
p_0^2 =   \vec p^2 + m^2 + \alpha_0^2 + \vec \delta^2 \pm 2 \sqrt{D_2(\vec p)},
\end{equation}
where
\begin{equation}
D_2(\vec p) = (\vec \delta \times \vec p)^2 + (\alpha_0 \vec p  -  m \vec \delta)^2 .
\end{equation}
Note that $D_2(\vec p) \ge 0$ as in the first case.

In both of these special cases, the dispersion relation is symmetric under
$p_0 \rightarrow - p_0$ indicating that positive and (reinterpreted) negative energy states are degenerate.
In addition, for a fixed value of $p_0$, the set of solutions for $\vec p$
forms a deformed sphere with two sheets where the radius as a function of angle 
is determined by the relevant factor, $D_1(\vec p)$, or $D_2(\vec p)$.
This simple geometric interpretation works well provided that the Lorentz-violation parameters
are small relative to the momentum and mass involved.  Special degeneracies may arise 
when the Lorentz-violating parameters become comparable to the size of the 
momentum or mass involved.

Recalling the relationship between the Lorentz violating parameters
and the quaternionic parameters
(\ref{alphadef3.1})-(\ref{epsilondef3.1}), we see that the
quaternionic representation determines mutually exclusive special
cases for which the analysis of the dispersion relation is easy.
These special cases partition the Lorentz violating parameters.
The case $\vec{\epsilon} = \vec{\delta}$, with corresponding condition $\vec \alpha=0$ 
corresponds to setting
$\hat b - \hat d_p = 0$ and $\hat{H}-\hat{G} = 0$, implying that $\hat b = \hat d_p =\hat H = \hat G =0$ since the relations must hold for arbitrary values of the momentum $\vec p$.  The other Lorentz-violation
parameters may be left arbitrary.
The case  $\vec{\epsilon} = -\vec{\delta}$, with corresponding condition $\alpha_0 =0$
corresponds to setting 
$\hat g - \hat h = 0$ and $b^0 +\hat{d}_1\cdot \hat{p} = 0$, while leaving the other parameters
arbitrary.  Again, this condition implies $\hat g = \hat h = b_0 = \hat d_1 =0 $ all vanish, while 
the other parameters remain arbitrary.

When the linear term in (\ref{disprel}) does not vanish, it is
sometimes still possible to give a complete analysis of the dispersion
relation.  We consider two special cases of particular interest: $b^\mu$, and $H^\mn$.

{\bf  Case:  $b^\mu \ne 0$, or $\hat{\epsilon} = -\hat{\delta}$}, but with $\alpha_0 \ne 0$:

This case corresponds to the special case where the only nonvanishing
Lorentz violating parameters are taken to be the vector $b^\mu.$  Using
the expression (\ref{disprel}), the dispersion relation becomes
\begin{equation}
(p_0^2 - \vec p^2 - m^2 - b_0^2 - \vec b^2)^2 - 
4 (b_0^2 \vec p^2 + b_0^2 \vec b^2 + (\vec b \cdot \vec p)^2 + m^2 \vec b^2)
+ 8 b_0 p_0 (\vec b \cdot \vec p) = 0 .
\end{equation}
Direct solution is difficult due to the nonvanishing linear coefficient in $p_0$,
however, we can still make some progress in obtaining the general structure
of the solution space by writing $\vec{b} = |\vec{b}| \hat{v}$ and decomposing $\vec{p}$
accordingly: $\vec{p} = t \hat{v} + \vec{z}$,
where $\vec z$ is the component of momentum perpendicular to $\hat v$.
Then 
\begin{equation}
(\vec{z}^2 - (p_0^2-m^2 + b_0^2 - \vec{b}^2 -t^2))^2 = 4\left(
(|\vec{b}|t -p_0b_0)^2 +m^2(\vec{b}^2-b_0^2)\right) .
\end{equation}
To solve for $|\vec{z}|$ as a function of $t,$ set 
\begin{eqnarray*}
f_1(t) & = & p_0^2-m^2 + b_0^2 - \vec{b}^2 -t^2 \\
f_2(t) & = & 4((|\vec{b}|t -p_0b_0)^2 +m^2(|\vec{b}|^2-b_0^2)).
\end{eqnarray*}
Note that for a solution to exist we must have $f_2(t) \geq 0$ and
$f_1(t) \geq -\sqrt{f_2(t)}.$  To proceed, assume $r^2 = b_0^2- \vec{b}^2
>0.$  Set 
\begin{eqnarray*}
x_0 & = &{1 \over r} (b_0p_0 - |\vec{b}|t) \\
x_1 & = & {1 \over r} (-|\vec{b}|p_0 + b_0t) 
\end{eqnarray*}
Then 
\begin{eqnarray*}
f_1(t) & = & x_0^2 - x_1^2 -m^2 + r^2  \\
f_2(t) & = & 4r^2(x_0^2 -m^2)
\end{eqnarray*}
and $f_1^2 -f_2$ factors:
\begin{equation}
f_1^2 -f_2 = ((x_1+r)^2 -(x_0^2 -m^2))((x_1-r)^2 -(x_0^2 -m^2)).
\end{equation}
When $\vec z=0$, the momentum points purely in the $\vec b$ direction
and $f_1^2 -f_2 = 0$ giving the endpoint conditions on the momentum:
\begin{eqnarray*}
(t+b_0)^2 & = & (p_0 +|\vec{b}|)^2 -m^2 \\
(t-b_0)^2 & = & (p_0 -|\vec{b}|)^2 -m^2 
\end{eqnarray*}
A similar analysis for the case $r = \vec b^2 - {b_0}^2
>0$ leads to the same pair of equations.
and thus, we generically obtain the four solutions:
\begin{eqnarray*}
t& = & - b_0 \pm \sqrt{(p_0 + |\vec{b}|)^2 -m^2}  \\
t& = &  b_0 \pm \sqrt{(p_0 - |\vec{b}|)^2 -m^2}.  
\end{eqnarray*}

{\bf  Case:  $H^{\mu \nu} \ne 0$, or $\hat{\epsilon} = \hat{\delta}$}, but with $\hat{\alpha} \ne 0$:

This case corresponds to the special case where the only nonvanishing
Lorentz violating parameters are taken to be the tensor $H^{\mu \nu}.$  Using
the expression (\ref{alphadef3.1})-(\ref{epsilondef3.1}), the coefficients in the dispersion
relation become
\begin{eqnarray*}
\beta_2 & = & -2[\vec{H}^2 + m^2 +\vec{p}^2 + \vec{h}^2] \\
\beta_1 & = & 8(\vec{H} \times \vec{p})\cdot \vec{h} \\
\beta_0 & = & \vec{p}^4 -2\vec{p}^2(\vec{H}^2 -m^2 +\vec{h}^2)
+(\vec{H}^2 +m^2)^2 -2m^2\vec{h}^2 -2\vec{h}^2\vec{H}^2 +\\
        &  &  \vec{h}^4 +4[(\vec{H}\cdot \vec{p})^2 + (\vec{h}\cdot
  \vec{p})^2 + (\vec{H}\cdot \vec{h})^2].
\end{eqnarray*}

To analyze the corresponding dispersion relation we begin by assuming
that $\vec{H}$ and $\vec{h}$ are colinear: $\vec{H} = s\vec{h}.$  Then 
\begin{eqnarray*}
\beta_2 & = & -2[ m^2 +\vec{p}^2 + (1+s^2)\vec{h}^2] \\
\beta_1 & = & 0 \\
\beta_0 & = & \vec{p}^4 -2\vec{p}^2((1+s^2)\vec{h}^2 -m^2)
+(s^2\vec{h}^2 +m^2)^2 -2m^2\vec{h}^2 -2s^2\vec{h}^4 +\\
        &  &  \vec{h}^4 +4[(1+s^2)(\vec{h}\cdot
  \vec{p})^2 + s^2\vec{h}^4].
\end{eqnarray*}
Completing the square, we obtain the solutions in a familiar form 
\beq
p_0^2 = \vec p^2 + m^2 + (1 + s^2) \vec h^2 \pm 2 \sqrt{D_3(\vec p)},
\eeq
where
\beq
D_3(\vec p) = (1+s^2) (\vec h \times  \vec p)^2 + m^2 \vec h^2.
\eeq	
Note that the variety (for fixed $p_0$) takes the form of two nested, pertubed spheres
in momentum space, as expected.

When $\vec{h}$ and $\vec{H}$ are not parallel, we can apply an observer Lorentz 
boost to reduce to
the parallel case.  To proceed, note that the components of $\vec{h}$
and $\vec{H}$ define the components of the antisymmetric 2-tensor
appearing in the Lagrangian (\ref{lag2.1})-(\ref{mass2.1}) via the
expression 
\begin{equation}
H^{\mu \nu}  =  \left( \begin{matrix}{
                0 & H^{1} & H^{2} & H^{3} \cr
               -H^{1} & 0 & h_3 & -h_2 \cr
               -H^{2} & -h_3 & 0 & h_1 \cr
               -H^{3} & h_2 & -h_1 & 0  }\end{matrix} \right) 
\end{equation}
In particular, the triple $(H^{\mu \nu}, \vec{H}, \vec{h})$ defines an
object which can be analyzed using the techniques used to treat the
triple $(F^{\mu \nu}, \vec{E},\vec{B})$ where $F^{\mu \nu}$ is
electromagnetic field strength, $\vec{E}$ is a (static) electric
field and $\vec{B}$ is a (static) magnetic field.  Assuming that
$|\vec{h}| > |\vec{H}|,$ write $\vec{H} = \vec{H}_{||} +
\vec{H}_{\perp}$ where $\vec{H}_{||} $ is the component of $\vec{H}$
in the direction of $\vec{h}$ and $\vec{H}_{\perp}$ is perpendicular
to $\vec{h}.$  Consider the boost 
\begin{equation}
\vec{u}  =  \frac{\vec{H}_{\perp} \times \vec{h}}{\vec{h}^2}
\end{equation}
and set $\gamma = (1-\vec{H}_{\perp}^2/\vec{h}^2)^{-\frac{1}{2}}.$
 Then, by direct calculation, in the coordinates associated to the boosted
frame $\vec{u}$ the vectors $\vec{H}$ and $\vec{h}$ are given by 
\begin{eqnarray*}
\vec{H}^\prime & = & \gamma \vec{H}_{||} \\
\vec{h}^\prime & = & \gamma^{-1}\vec{h}.
\end{eqnarray*}
A similar argument applies when $|\vec h| < |\vec H|$ by interchanging the roles
of the vectors and  the analysis is reduced to the previous special case
in the new frame.   

\vglue 0.6cm
{\bf \noindent V. SUMMARY}
\vglue 0.4cm

In this paper we have provided a detailed investigation of the
dispersion relation associated to the Dirac operator for the SME using
an approach which employed quaternions.  The traditional $4\times 4$ structure
of the Dirac equation can be re-expressed in $2 \times 2$ form where
the entries are quaternion valued.  This greatly simplifies the matrix
structure at the expense of loss of commutativity of the matrix
elements.  While it is possible to perform all of the computations
using the traditional $4\times 4$ notation, the quaternion valued  
elements provide an efficient way to organize the computation.

Generically, our results exhibit the dispersion relation as a pair of
perturbed spheres.  We find that the dispersion relation can be easily
solved for two special sets of parameter choices.  Each choice allows half
of the Lorentz violating parameter space to be "turned on" while the
other half of the space is "turned off".  
This provides a practical way to approach analysis involving the fermions in the
SME with each special case providing simple insight into the effect of
the relevant terms.  Many problems that are intractable using a
general fourth-order dispersion relation become far simpler to analyze
when the dispersion relation factors into two  second-order ones.  In
particular, the analysis of the dispersion relation for a theory
involving Lorentz violation coupled to spin 
through the $H^{\mu \nu}$ term in the Lagrangian defined by
(\ref{lag2.1})-(\ref{mass2.1}) can be carried out explicitly.  In
addition to the above results, further exploitation of the representation (for
example, the use of the dispersion relation to label eigenstates)
provide interesting directions for probing hidden symmetries of the
SME. 

\vglue 0.6cm
{\bf \noindent ACKNOWLEDGMENTS}
We wish to acknowledge the support of New College of Florida's faculty
development funds that contributed to the successful completion of
this project.

\vglue 0.6cm
{\bf\noindent REFERENCES}
\vglue 0.4cm


\begin{thebibliography}{xx}

\bibitem{kps}
V.A.\ Kosteleck\'y and S.\ Samuel,
Phys.\ Rev.\ D {\bf 39}, 683 (1989);
{\it ibid.} 
{\bf 40}, 1886 (1989);
Phys.\ Rev.\ Lett.\ {\bf 63}, 224 (1989);
{\it ibid.} 
{\bf 66}, 1811 (1991);
V.A.\ Kosteleck\'y and R.\ Potting,
Nucl.\ Phys.\ B {\bf 359}, 545 (1991);
Phys.\ Lett.\ B {\bf 381}, 89 (1996);
Phys.\ Rev.\ D {\bf 63}, 046007 (2001); 
V.A.\ Kosteleck\'y, M.\ Perry, and R.\ Potting,
Phys.\ Rev.\ Lett.\ {\bf 84}, 4541 (2000). 

\bibitem{ck} 
D.\ Colladay and V.A.\ Kosteleck\'y,
Phys.\ Rev.\ D {\bf 55}, 6760 (1997);
Phys.\ Rev.\ D {\bf 58}, 116002 (1998).

\bibitem{cpt98}
For a summary of recent theoretical models and
experimental tests
see, for example,
{\it CPT and Lorentz Symmetry III}, V.A.\ Kosteleck\'y, ed., 
World Scientific, Singapore, 2005; 
{\it CPT and Lorentz Symmetry IV}, V.A.\ Kosteleck\'y, ed.,
World Scientific, Singapore, 2008.

\bibitem{kle} 
V.A.\ Kosteleck\'y and R.\ Lehnert,
Phys.\ Rev.\ D {\bf 63}, 065008 (2001).

\bibitem{klp1}
V.A.\ Kosteleck\'y, C.\ Lane, and A.\ Pickering,
Phys.\ Rev.\ D {\bf 65}, 056006 (2002).

\bibitem{klp2}
V.A.\ Kosteleck\'y and A.\ Pickering,
Phys.\ Rev.\ Lett.\ {\bf 91}, 031801 (2003).

\bibitem{cm1}
D.\ Colladay and P.\ McDonald,
Phys.\ Rev.\ D {\bf 75}, 105002 (2007);
Phys.\ Rev.\ D {\bf 77}, 085006 (2008);
Phys.\ Rev.\ D {\bf 79}, 125019 (2009).

\bibitem{kpott}
A.\ Kosteleck\'y and R.\ Potting,
Phys\ Rev.\ D {\bf 79}, 065018 (2009).

\bibitem{kmewes}
A.\ Kostelecky and M.\ Mewes,
Phys.\ Rev.\ Lett.\ {\bf 99}, 011601 (2007).

\bibitem{neil}
V.\ A.\ Kosteleck\'y and N.\ Russell
arXiv:0801.0287

\bibitem{ralph2}
R.\ Lehnert,
J.\ Math.\ Phys.\ {\bf 45} 3399 (2004).

\bibitem{quatdirac}
See, for example,
D.\ Schuricht and M.\ Greiter
Eur.\ J.\ Phys.\ {\bf 25} 755 (2004).

\bibitem{colmac2}
D.\ Colladay and P.\ McDonald,
J.\ Math.\ Phys.\ {\bf 43} 3554 (2002).

\bibitem{ba}
B.\ Altschul,
J.\ Phys.\ A {\bf 39},13757 (2006).

\bibitem{ralf}
 R. Lehnert,
Phys.\ Rev.\ D {\bf 63}, 065008 (2001).

\end{thebibliography}
\end{document}